# Classifying economics for the common good:

# Connecting sustainable development goals to JEL codes

Jussi Heikkilä[1]

*April 2020 version*

**Abstract**

How does economics research help in solving societal challenges? This brief note sheds additional light on this question by providing ways to connect Journal of Economic Literature (JEL) codes and Sustainable Development Goals (SDGs) of the United Nations. These simple linkages illustrate that the themes of SDGs have corresponding JEL classification codes. As the mappings presented here are necessarily imperfect and incomplete, there is plenty of room for improvements. In an ideal world, there would be a JEL classification system for SDGs, a separate JEL code for each of the 17 SDGs.

[1] email: jussi.heikkila@jyu.fi I thank Salla Laukkanen for excellent research assistance.

# 1 Introduction

Economists play an increasingly important role in helping governments design new policies and regulations (Duflo 2017) but how does academic economics research help in solving global challenges? One way to analyse this is to explore the link between economic research and the Sustainable Development Goals (SDGs, LaFleur 2019). In 2015, the member states of the United Nations adopted 17 sustainable development goals, as part of the 2030 Agenda for Sustainable Development (United Nations 2015a). Great progress has been made but there is still a lot of work to do before achieving these goals (United Nations 2019).[2] Indexing and classification of documents are important since they decreases search costs and therefore promote efficient allocation of attention. In the context of economics research, Journal of Economic Literature (JEL) codes have been used to classify economic research since the early 20th century (Cherrier, 2019). However, there does not yet exist an explicit JEL coding for sustainable development goals.

This brief note suggests that the existing classification systems of economic literature could be utilized to track how much attention is allocated within economics research to topics that are related to global challenges. The contribution of this brief note to existing literature is to connect SDGs to JEL codes. The remainder of the paper is organized as follows. Section 2 presents SDGs and JEL classification system. Section 3 links JEL codes and SDGs goal by goal and discusses the limitations. Section 4 concludes.

# 2 SDGs and JEL codes

## 2.1 SDGs

According to the United Nations, sustainable development is defined as "development that meets the needs of the present without compromising the ability of future generations to meet their own needs".[3] The 17 sustainable development goals succeeded the United Nations' eight Millennium development goals (MDGs) in 2015 (United Nations 2015a, 2015b).[4] The eight MDGs were international development goals for the year 2015 established following the Millennium Summit of the United Nations and adoption of the United Nations Millennium Declaration in September 2000.[5] In September 2015, 193 countries of the UN General Assembly adopted the 2030 Development Agenda ("Agenda 2030") which paragraph 59 outlines the 17 Sustainable Development Goals and the associated 169 targets and 232 indicators (United Nations 2015a). Currently, the aim is to achieve all SDGs by 2030 and the progress is monitored continuously (United Nations 2015, 2019; Sachs 2019). The 17 SDGs are:

---

[2] See e.g. SDG Tracker of Our World In data: https://sdg-tracker.org/ (Accessed on 21 March 2020), Sustainable Development Report: https://www.sdgindex.org/ (Accessed on 21 March 2020) and Sachs et al. (2019).

[3] See https://www.un.org/sustainabledevelopment/development-agenda/ Accessed on 21 Mar 2020.

[4] The MDGs were: Goal 1. Eradicate extreme poverty and hunger; Goal 2. Achieve universal primary education; Goal 3. Promote gender equality and empower women; Goal 4. Reduce child mortality; Goal 5. Improve maternal health; Goal 6. Combating HIV/AIDS, malaria, and other diseases; Goal 7. Ensure environmental sustainability; Goal 8. Develop a global partnership for development. See Millennium development goals: https://www.un.org/millenniumgoals/ SDGs: https://www.un.org/sustainabledevelopment/sustainable-development-goals/ Accessed on 21 March 2020.

[5] See https://www.un.org/millenniumgoals/bkgd.shtml Accessed on 4 April 2020

Goal 1. End poverty in all its forms everywhere
Goal 2. End hunger, achieve food security and improved nutrition and promote sustainable agriculture
Goal 3. Ensure healthy lives and promote well-being for all at all ages
Goal 4. Ensure inclusive and equitable quality education and promote lifelong learning opportunities for all
Goal 5. Achieve gender equality and empower all women and girls
Goal 6. Ensure availability and sustainable management of water and sanitation for all
Goal 7. Ensure access to affordable, reliable, sustainable and modern energy for all
Goal 8. Promote sustained, inclusive and sustainable economic growth, full and productive employment and decent work for all
Goal 9. Build resilient infrastructure, promote inclusive and sustainable industrialization and foster innovation
Goal 10. Reduce inequality within and among countries
Goal 11. Make cities and human settlements inclusive, safe, resilient and sustainable
Goal 12. Ensure sustainable consumption and production patterns
Goal 13. Take urgent action to combat climate change and its impacts
Goal 14. Conserve and sustainably use the oceans, seas and marine resources for sustainable development
Goal 15. Protect, restore and promote sustainable use of terrestrial ecosystems, sustainably manage forests, combat desertification, and halt and reverse land degradation and halt biodiversity loss
Goal 16. Promote peaceful and inclusive societies for sustainable development, provide access to justice for all and build effective, accountable and inclusive institutions at all levels
Goal 17. Strengthen the means of implementation and revitalize the global partnership for sustainable development

In order to get a rough overview of the occurrence of MDGs and SDGs in economics research during the past two decades, we conduct some very general keyword searches. Figure 2 shows the trends of simple keyword searches in IDEAS/RePEc bibliographic database[6] for both search queries "sustainable development goal" and "millennium development goal". As expected, the trends in the number of publications clearly show that SDGs have replaced MDGs after Agenda 2030 was introduced in 2015. It is also noteworthy that the increase in the number of research papers mentioning SDGs is growing very steeply.

[6] IDEAS is an economics-focused bibliographic database with over 3,100,000 indexed items of research, see https://ideas.repec.org/. It is based on RePEc database, see http://repec.org/. Accessed on 4 April 2020.

**Figure 1. Publications mentioning MDGs or SDGs**

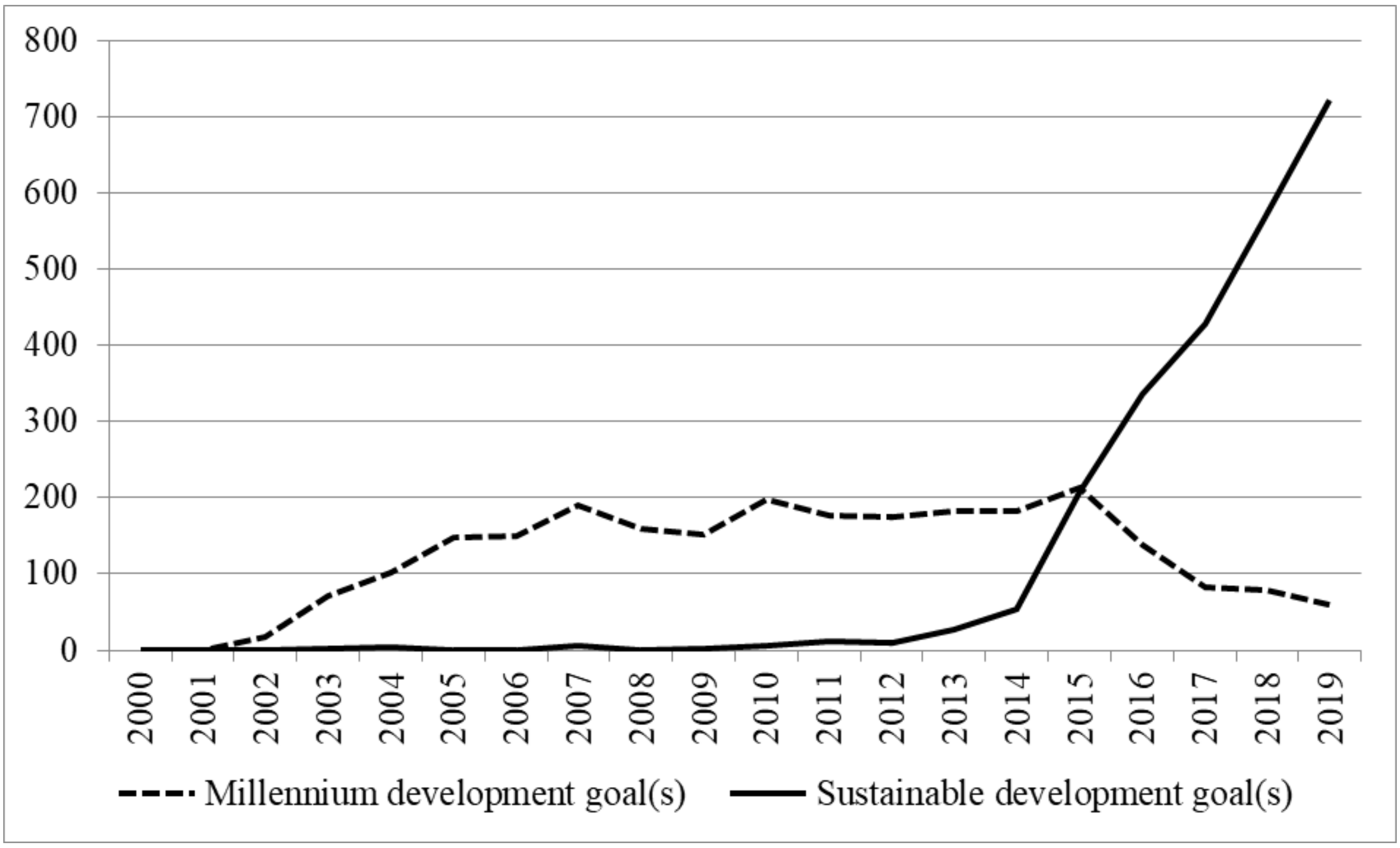

Notes: Source is IDEAS/RePEc search, available at https://ideas.repec.org/search.html. Search queries: SDGs "sustainable development goal" OR "sustainable development goals" and MDGs "millennium development goal" OR "millennium development goals". No filters applied, i.e., search "All" (incl. Articles, Papers, Chapters, Books, Software) in "Whole record". Retrieved on 4 April 2020.

## 2.2 JEL codes

In the field of economics, Journal of Economic Literature classification is the *de facto* standard in classifying articles (Cherrier 2017; Kosnik 2018). Majority of economics journals require authors to choose JEL codes which best describe their manuscripts when submitting (Kosnik 2018) and bibliographic databases such as IDEAS and SSRN allow users to browse articles by JEL codes. The first version of the JEL codes (or EconLit subject descriptors) by the American Economic Association (AEA) was published in 1911 and it has since been occasionally updated and extended (Cherrier 2017). According to the webpage of the AEA:

*"The JEL classification system was developed for use in the Journal of Economic Literature (JEL), and is a standard method of classifying scholarly literature in the field of economics. The system is used to classify articles, dissertations, books, book reviews, and working papers in EconLit, and in many other applications."*[7]

As of the beginning of 2020, there are 20 JEL categories, which have 122 second level sub-categories that have altogether 856 third level sub-categories. Table 1 shows that the number of sub-categories varies across main JEL classes. Interestingly, despite the important role of JEL codes, they have received only limited attention in research until recently (Cherrier 2017; Kosnik 2018).

[7] https://www.aeaweb.org/econlit/jelCodes.php?view=jel Accessed on 21 March 2020.

**Table 1. JEL codes**

| | 1st level JEL codes | Number of 2nd level codes | Number of 3rd level codes |
|---|---|---|---|
| 1 | A General Economics and Teaching | 3 | 16 |
| 2 | B History of Economic Thought, Methodology, and Heterodox Approaches | 5 | 32 |
| 3 | C Mathematical and Quantitative Methods | 9 | 70 |
| 4 | D Microeconomics | 9 | 65 |
| 5 | E Macroeconomics and Monetary Economics | 7 | 47 |
| 6 | F International Economics | 6 | 53 |
| 7 | G Financial Economics | 5 | 33 |
| 8 | H Public Economics | 8 | 56 |
| 9 | I Health, Education, and Welfare | 3 | 23 |
| 10 | J Labor and Demographic Economics | 8 | 62 |
| 11 | K Law and Economics | 4 | 30 |
| 12 | L Industrial Organization | 9 | 72 |
| 13 | M Business Administration and Business Economics, Marketing, Accounting, Personnel Economics | 5 | 29 |
| 14 | N Economic History | 9 | 74 |
| 15 | O Economic Development, Innovation, Technological Change, and Growth | 5 | 41 |
| 16 | P Economic Systems | 5 | 43 |
| 17 | Q Agricultural and Natural Resource Economics, Environmental and Ecological Economics | 5 | 49 |
| 18 | R Urban, Rural, Regional, Real Estate, and Transportation Economics | 5 | 31 |
| 19 | Y Miscellaneous Categories | 9 | 11 |
| 20 | Z Other Special Topic | 3 | 19 |
| | Min | 3 | 11 |
| | Max | 9 | 74 |
| | Average | 6.10 | 42.80 |
| | Median | 5 | 42 |
| | Total sum | 122 | 856 |

Notes: "General" JEL classes under 1st level codes with no 2nd level codes (e.g., B00 and C00) have been counted as 3rd level codes.

# 3 Connecting SDGs and JEL codes

## 3.1 Methods and data

In an ideal or the first best world, there would be own JEL classification for SDGs — that is, a separate JEL code for each of the 17 SDGs. However, currently we live in a second best world without this type of classification.[8] In order to fill in this gap, we provide three simple ways to link

[8] Note that JEL code "Q01 Sustainable Development" "covers studies about issues related to sustainable development in the broadest context. That is, studies should include agriculture, natural resources, energy sources, and the environment."

JEL codes and SDGs. In principle, we could either link "each JEL code to the closest SDGs" or "each SDG to the closest JEL codes". We chose the latter option that is much simpler and link 17 SDGs to the most similar 3rd level JEL codes. JEL code data was collected from the AEA's webpage and SDGs were collected from United Nations' webpage (see Section 2). Three alternative methods to establish links between SDGs and JEL codes are described next.

First, we extract keywords directly from the 17 SDGs (Section 2.1) and search over the keywords of 856 3rd level JEL codes (Table 1) to document the number of potentially related JEL codes. For instance, the keyword for Goal 1 "End poverty in all its forms everywhere" is "poverty". Some SDGs are more broadly defined and are assigned more keywords correspondingly: For instance, the keywords for Goal 9 "Build resilient infrastructure, promote inclusive and sustainable industrialization and foster innovation" are "infrastructure", "industrialization" and "innovation". We provide examples of JEL codes that most closely relate to the selected keywords.

Second, we build on LaFleur (2019) who created an SDG classification system for DESA publications. LaFleur (2019) was among the first to analyze United Nations publications using machine-learning approach to compute how much each SDG is represented in individual publications.[9] As part of his analysis, LaFleur (2019) used machine learning methods to identify keywords related to each SDG and we use a slightly modified version of his keyword list as a basis to match SDGs to JELs. For each SDG, we rank JEL codes according to the number of overlapping keywords. These keywords are presented in Table 2. We explore three different weightings for the keywords: 1) no weighting, 2) weight 1/1 for the 1st keyword, 1/2 for the 2nd, 1/3 for the 3rd and so forth, and 3) weight 1 for the first five keywords and 1/[number of order] for the rest.

Third, we select three most relevant keywords for each SDG from the LaFleur's (2019) list based on elimination and discretion. First, we delete "general words" that clearly are not related solely to a specific SDG such as "change", "impacts", "patterns", "rapid", "added", "policies", "capita" and "nations". Second, we eliminate pairs of words when the same word-pair occurs also separately in the list. Third, in most cases we delete plural forms if the same word is on the list also as singular. Finally, from the remaining set of words, we select three words that most closely relate to corresponding SDG based on discretion. For most SDGs, we are left with three singular nouns as shown in Table 2.

[9] See also Körfgen et al. (2018).

**Table 2. Keywords for SDGs**

| SDG | LaFleur (2019) keywords | Selected three |
|---|---|---|
| 1 | poverty social protection poor social_protection extreme disasters extreme_poverty poverty_line day end_pov-erty line losses living disaster cash protection_systems person poor_vulnerable disaster_risk | poverty poor social_protection |
| 2 | food hunger agricultural agriculture children malnutrition production genetic prices export markets hungry food_production subsidies food_security nutrition undernourished breeds aid insecurity | food agriculture nutrition |
| 3 | health deaths diseases people mortality births live_births maternal children hiv age live care rate years_age reproductive deaths_live worldwide risk women | health disease mortality |
| 4 | education primary children school primary_education quality secondary schools learning quality_education skills secondary_education reading primary_school proficiency mathematics teachers minimum saharan basic | education school learning |
| 5 | women girls gender women_girls equality gender_equality violence sexual age marriage female married work rights genital_mutilation mutilation female_genital genital partner globally | women gender equality |
| 6 | water sanitation management water_sanitation people drinking_water drinking population improved hygiene facilities safely water_resources global_population wastewater freshwater water_scarcity scarcity resources water_stress | water sanitation hygiene |
| 7 | energy electricity renewable renewable_energy clean affordable modern cooking fuels access energy_efficiency energy_consumption energy_intensity intensity consumption access_electricity reliable efficiency technologies affordable_reliable | energy electricity renewable |
| 8 | growth labour employment unemployment work decent financial productivity decent_work financial_services productive men adults working youth economic_growth child labour_productivity jobs developed | labor employment productivity |
| 9 | manufacturing infrastructure developing added manufacturing_added industrialization innovation gdp devel-oped employment industries industrial mobile research job research_development intensity resilient_infrastruc-ture resilient emissions | infrastructure industrialization innovation |
| 10 | inequality developed income developing duty exports oda money duty_free developing_states tariff remit-tances migration treatment reducing inequalities products migrant policies island | inequality income migration |
| 11 | cities urban waste air pollution slums urban_population solid_waste solid land urbanization management disas-ters air_pollution rapid safe resilient housing inclusive risk | cities urban housing |
| 12 | consumption production material consumption_production sustainable sustainable_consumption water conven-tion material_consumption food domestic_material domestic impacts patterns natural capita production_pat-terns environmental pollutants wastes | consumption production sustainable |
| 13 | climate change climate_change agreement paris paris_agreement action global parties emissions adaptation convention temperature framework nations framework_convention united_nations determined climate_action degrees | climate climate_change emission |
| 14 | marine oceans ocean coastal resources areas fisheries ecosystems pollution protected_areas marine_resources fish overfishing biodiversity protected management stocks eutrophication ocean_acidification acidification | marine ocean fish |
| 15 | biodiversity land species forests areas loss forest degradation wildlife desertification protected ecosystems ter-restrial conservation resources halt land_degradation management covered biodiversity_loss | biodiversity land forest |
| 16 | institutions rights justice violence inclusive victims access_justice children human_rights data human societies trafficking effective peaceful levels sexual forms birth_registration registration | institutions rights justice |
| 17 | developed development data oda developing capacity registration partnerships capacity_building building regions trade received agenda statistical enhance debt areas complete death_registration | development official_development_assistance trade |

## 3.2 Results

Table 3 reports the results of the simple keyword search approach. We also provide an example of a relevant JEL code for each keyword if there is one. This very simple keyword search already indicates that economic research classification can be relatively easily linked to SDGs. Only for some keywords no matching JEL codes are found (e.g., “girl”, “sanitation”).

**Table 3. Link between SDGs and JEL codes, simple keyword search**

| Sustainable Development Goal | Keywords | Number of 3rd level JEL codes | Example |
|---|---|---|---|
| Goal 1. End poverty in all its forms everywhere | Poverty | 9 | I32 Measurement and Analysis of Poverty |
| Goal 2. End hunger, achieve food security and improved nutrition and promote sustainable agriculture | Hunger | 1 | O15 Human Resources • Human Development • Income Distribution • Migration |
| | Food | 12 | Q18 Agricultural Policy • Food Policy |
| | Nutrition | 4 | I12 Health Behavior |
| | Agricultur | 17 | O13 Agriculture • Natural Resources • Energy • Environment • Other Primary Products |
| Goal 3. Ensure healthy lives and promote well-being for all at all ages | Health | 21 | I15 Health and Economic Development |
| | Well-being | 3 | I31 General Welfare, Well-Being |
| Goal 4. Ensure inclusive and equitable quality education and promote lifelong learning opportunities for all | Education | 20 | I25 Education and Economic Development |
| | Learning | 5 | J24 Human Capital • Skills • Occupational Choice • Labor Productivity |
| Goal 5. Achieve gender equality and empower all women and girls | Gender | 7 | K38 Human Rights Law • Gender Law |
| | Women | 2 | J16 Economics of Gender • Non-labor Discrimination |
| | Girl | 0 | - |
| Goal 6. Ensure availability and sustainable management of water and sanitation for all | Water | 6 | Q25 Water |
| | Sanitation | 0 | - |
| Goal 7. Ensure access to affordable, reliable, sustainable and modern energy for all | Energy | 12 | O13 Agriculture • Natural Resources • Energy • Environment • Other Primary Products |
| Goal 8. Promote sustained, inclusive and sustainable economic growth, full and productive employment and decent work for all | Economic growth | 5 | O47 Empirical Studies of Economic Growth • Aggregate Productivity • Cross-Country Output Convergence |
| | Employment | 24 | E24 Employment • Unemployment • Wages • Intergenerational Income Distribution • Aggregate Human Capital • Aggregate Labor Productivity |
| | Work | 29 | J81 Working Conditions |
| Goal 9. Build resilient infrastructure, promote inclusive and sustainable industrialization and foster innovation | Infrastructure | 9 | H54 Infrastructures • Other Public Investment and Capital Stock |
| | Industrialization | 3 | L52 Industrial Policy • Sectoral Planning Methods |
| | Innovation | 7 | O32 Management of Technological Innovation and R&D |
| Goal 10. Reduce inequality within and among countries | Inequality | 12 | J15 Economics of Minorities, Races, Indigenous Peoples, and Immigrants • Non-labor Discrimination |
| Goal 11. Make cities and human settlements inclusive, safe, resilient and sustainable | Cities | 6 | R23 Regional Migration • Regional Labor Markets • Population • Neighborhood Characteristics |
| | Housing | 11 | R31 Housing Supply and Markets |
| | Transport | 18 | O18 Urban, Rural, Regional, and Transportation Analysis • Housing • Infrastructure |
| Goal 12. Ensure sustainable consumption and production patterns | Consumption | 11 | E21 Consumption • Saving • Wealth |
| | Production | 28 | D62 Externalities |
| Goal 13. Take urgent action to combat climate change and its impacts* | Climate | 3 | Q58 Government Policy |
| | Climate change | 1 | Q54 Climate • Natural Disasters and Their Management • Global Warming |
| Goal 14. Conserve and sustainably use the oceans, seas and marine resources for sustainable development | Ocean | 2 | Q25 Water |
| | Marine or maritime | 2 | Q22 Fishery • Aquaculture |
| Goal 15. Protect, restore and promote sustainable use of terrestrial ecosystems, sustainably manage forests, combat desertification, and halt and reverse land degradation and halt biodiversity loss | Ecosystem | 2 | Q57 Ecological Economics: Ecosystem Services • Biodiversity Conservation • Bioeconomics • Industrial Ecology |
| | Forest | 3 | Q23 Forestry |
| | Desertification | 3 | O13 Agriculture • Natural Resources • Energy • Environment • Other Primary Products |
| | Land | 17 | Q24 Land |
| | Biodiversity | 1 | Q57 Ecological Economics: Ecosystem Services • Biodiversity Conservation • Bioeconomics • Industrial Ecology |
| Goal 16. Promote peaceful and inclusive societies for sustainable development, provide access to justice for all and build effective, accountable and inclusive | Peace | 2 | D74 Conflict • Conflict Resolution • Alliances • Revolutions |
| | Justice | 3 | D63 Equity, Justice, Inequality, and Other Normative Criteria and Measurement |
| | Institution | 22 | D02 Institutions: Design, Formation, Operations, and Impact |
| Goal 17. Strengthen the means of implementation and revitalize the global partnership for sustainable development | International | 63 | F02 International Economic Order and Integration |

Table 4 lists for each SDG three 3rd level JEL codes that have the most overlap with keywords suggested by LaFleur (2019). We explored three different weightings based on the ranking of keywords and they produced relatively similar rankings for JEL codes. Table 4 reports the results for the third weighting option (see Section 3.1). For most SDGs, there are JEL codes that clearly have most keyword overlap with LaFleur's keywords, whereas for some the best matching JEL codes are not so unambiguous. In other words, there are several JEL codes with equal amount of keyword overlap. Most of the identified JEL codes seem to match SDGs quite well. One example of an exception is Goal 13 where JEL code C22 "Time-series models" overlaps in keywords but is not actually related specifically to the substance of Goal 13.

Table 5 is otherwise similar to Table 4 except that there the number of keywords for each SDG is limited to three most relevant ones (see Table 2). The resulting JEL codes are in most cases the same as in Table 4 but for some there is less overlap. For instance, for SDG 17 the top three JEL codes are all different in Tables 4 and 5. To summarize, these very simple keyword similarity analyses suggest that it is possible to connect SDGs and JEL codes. Thus, economists may classify their research articles to JEL codes that are closely linked to SDGs.

## Table 4. Link between SDGs and JEL codes, modified keywords of LaFleur (2019)

| SDG | Top three JEL codes | | Covers studies about/related to | Overlapping keywords |
|---|---|---|---|---|
| Goal 1. End poverty in all its forms everywhere | I32 | Measurement and Analysis of Poverty | issues related to poverty, its measurement and analysis. | poverty, poor, poverty line, line |
| | O15 | Human Resources • Human Development • Income Distribution • Migration | issues related to labor, demography, health, education, and welfare in the context of economic development and developing economies. Includes studies on nutrition, health, education, fertility, household structure and formation, labor markets and social policy. | poverty, social, living |
| | H53* | Government Expenditures and Welfare Programs | issues related to government expenditures on welfare and related policies, including food stamp programs and studies about HUD. | poverty, social |
| Goal 2. End hunger, achieve food security and improved nutrition and promote sustainable agriculture | O13 | Agriculture • Natural Resources • Energy • Environment • Other Primary Products | development issues related to agriculture, natural resources, other primary products, energy, and the environment. | food, agricultural, agriculture, production, food production, markets |
| | Q11 | Aggregate Supply and Demand Analysis • Prices | issues related to aggregate agricultural market, including supply, demand, prices and sustainable agriculture. | food, agricultural, agriculture, production, food production |
| | Q13 | Agricultural Markets and Marketing • Cooperatives • Agribusiness | issues related to agricultural markets and marketing, cooperatives, and agribusiness. | food, agricultural, agriculture |
| Goal 3. Ensure healthy lives and promote well-being for all at all ages | I12 | Health Behavior | issues related to health production and conditions and their consequences. Covers studies about the causes and effects of being healthy or unhealthy, including studies about quality-adjusted life years. Includes studies on nutrition, mortality, morbidity, suicide, substance abuse and addiction, and disability as related to economic behavior. Studies should be cross-classified here and under I11 if they cover topics in both categories. | health, deaths, mortality, rate |
| | I14 | Health and Inequality | the impact of health on social and economic inequality, and the impact of inequality on health. May cross-list with O15. | deaths, mortality, births, age, rate |
| | I15* | Health and Economic Development | the impact of health on economic development, and the impact of levels of development on health. May cross-list with O15. | health, mortality |
| Goal 4. Ensure inclusive and equitable quality education and promote lifelong learning opportunities for all | I21 | Analysis of Education | all the economic issues, including output, quality, and demand and supply, related to education, except for educational finance. | education, primary, school, quality, secondary, learning, primary school |
| | A21 | Pre-college | issues related to teaching economics at a pre-college level. For example, it includes studies on how to teach economics, what to teach in economics, and the outcomes from economic education at the pre-college level. Also covers the economic knowledge of the public at large. | education, primary, school |
| | I24 | Education and Inequality | the impact of education on social and economic inequality, and the impact of inequality on education. May cross-list with O15. | education, school, quality |
| Goal 5. Achieve gender equality and empower all women and girls | J16 | Economics of Gender • Non-labor Discrimination | economic issues related to gender, except for labor market discrimination, which is classified under J71 or J78. | women, gender, equality, gender equality, sexual, female |
| | J71 | Discrimination | issues related to labor discrimination, including wage discrimination and discrimination in hiring and firing. | women, gender, equality, sexual, age |
| | I24* | Education and Inequality | the impact of education on social and economic inequality, and the impact of inequality on education. May cross-list with O15. | gender, equality |
| Goal 6. Ensure availability and sustainable management of water and sanitation for all | Q53 | Air Pollution • Water Pollution • Noise • Hazardous Waste • Solid Waste • Recycling | issues related to environmental degradation, its impacts and its solution. | water, management |
| | R53 | Public Facility Location Analysis • Public Investment and Capital Stock | issues related to public facility location and public investment, including infrastructure. | management, facilities |
| | Q25 | Water | issues related to water as a renewable resource and its conservation. | water, wastewater |
| Goal 7. Ensure access to affordable, reliable, sustainable and modern energy for all | Q42 | Alternative Energy Sources | issues related to alternative energy sources, for example non-hydrocarbon fuels or wind. | energy, renewable, renewable energy |
| | L94 | Electric Utilities | issues related to the electric utility industry. | enertgy, electricity |
| | Q48 | Government Policy | issues related to government policy on energy. | energy, clean |
| Goal 8. Promote sustained, inclusive and sustainable economic growth, full and productive employment and decent work for all | E24 | Employment • Unemployment • Wages • Intergenerational Income Distribution • Aggregate Human Capital • Aggregate Labor Productivity | issues related to the aggregate labor market, including unemployment, wage level, wage indexation, intergenerational income distribution and aggregate labor productivity. Also covers studies about intergenerational income mobility. | employment, unemployment, work, productivity, men |
| | E31 | Price Level • Inflation • Deflation | issues related to aggregate price levels, inflation, and deflation, including the CPI and the Phillips curve. | growth, employment, unemployment, men |
| | J68 | Public Policy | public policy issues related to labor mobility and unemployment, including employment services. | employment, unemployment, work, men |
| Goal 9. Build resilient infrastructure, promote inclusive and sustainable industrialization and foster innovation | O14 | Industrialization • Manufacturing and Service Industries • Choice of Technology | issues related to industrialization and industries in developing countries, including manufacturing, service and transportation and the choice of technology. | manufacturing, industrialization, industrial |
| | O11 | Macroeconomic Analyses of Economic Development | macroeconomic issues related to development, including economic growth of developing countries. These studies include both theoretical models and empirical studies (time series and cross-sectional). | developing, gdp |
| | F63 | Economic Development | the impact of globalization on economic development, including poverty, labor markets, and gender issues. | developing, developed |

## Table 4. Continued

| | | | | |
|---|---|---|---|---|
| Goal 10. Reduce inequality within and among countries | F63 | Economic Development | the impact of globalization on economic development, including poverty, labor markets, and gender issues. | inequality, developed, developing |
| | F61* | Microeconomic Impacts | to the impact of globalization on various microeconomic issues, including income distribution, market structure and pricing, and the behavior of consumers and firms. | inequality, income |
| | I32* | Measurement and Analysis of Poverty | issues related to poverty, its measurement and analysis. | inequality, income |
| Goal 11. Make cities and human settlements inclusive, safe, resilient and sustainable | Q53 | Air Pollution • Water Pollution • Noise • Hazardous Waste • Solid Waste • Recycling | issues related to environmental degradation, its impacts and its solution. | waste, air, pollution, solid waste, solid, land, management, air pollution |
| | K32 | Energy, Environmental, Health, and Safety Law | the intersections of environmental, health and safety laws and economics or economies. | waste, air, pollution, management, safe |
| | R11 | Regional Economic Activity: Growth, Development, Environmental Issues, and Changes | issues related to regional economic activity including growth, development, and changes. Studies may also include the effects of agglomeration economies and foreign linkages (for example foreign direct investment or foreign trade) on regional economic activities. Includes environmental aspects of regional economics, such as the regional effects of climate change. | cities, urban, slums, urbanization |
| Goal 12. Ensure sustainable consumption and production patterns | Q11 | Aggregate Supply and Demand Analysis • Prices | issues related to aggregate agricultural market, including supply, demand, prices and sustainable agriculture. | productiion, sustainable, food |
| | E27* | Forecasting and Simulation: Models and Applications | issues related to forecasting or simulation of aggregate consumption, saving, production, employment and/or investment. | consumption, production |
| | D62* | Externalities | mostly theoretical studies about issues related to externalities and welfare analysis. | consumption, production |
| Goal 13. Take urgent action to combat climate change and its impacts* | Q54 | Climate • Natural Disasters and Their Management • Global Warming | issues related to climate and natural disasters, including desertification and drought. | climate, change, climate change, global, emissinos |
| | C22 | Time-Series Models • Dynamic Quantile Regressions • Dynamic Treatment Effect Models • Diffusion Processes | econometric issues related to a single time series variable or single-equation models using time-series variables. These studies include such subjects as: unit root tests, trend or difference stationarity, autocorrelated error terms, AR(I)MA models, switching regression models, single-equation error correction models, and spectral analysis. | change, action |
| | J52 | Dispute Resolution: Strikes, Arbitration, and Mediation • Collective Bargaining | issues related to labor dispute resolution, including collective bargaining. | agreement, action |
| Goal 14. Conserve and sustainably use the oceans, seas and marine resources for sustainable development | Q25 | Water | issues related to water as a renewable resource and its conservation. | oceans, ocean, pollution |
| | L92 | Railroads and Other Surface Transportation | issues related to surface transportation, including trains, autos, buses, trucks, and water carries. Also covers studies related to ports as they influence the performance of the transportation industry. | marine, ocean |
| | K32 | Energy, Environmental, Health, and Safety Law | the intersections of environmental, health and safety laws and economics or economies. | resources, pollution, management |
| Goal 15. Protect, restore and promote sustainable use of terrestrial ecosystems, sustainably manage forests, combat desertification, and halt and reverse land degradation and halt biodiversity loss | Q57 | Ecological Economics: Ecosystem Services • Biodiversity Conservation • Bioeconomics • Industrial Ecology | issues related to topics in ecological economics including ecosystem services and biodiversity. | biodiversity, species, wildlife, conservation, management |
| | O13 | Agriculture • Natural Resources • Energy • Environment • Other Primary Products | development issues related to agriculture, natural resources, other primary products, energy, and the environment. | land, forest, desertification, resources, management |
| | Q24 | Land | issues related to land as a renewable resource and its conservation. | land, degradation, desertification, conservation |
| Goal 16. Promote peaceful and inclusive societies for sustainable development, provide access to justice for all and build effective, accountable and inclusive institutions at all levels | O17 | Formal and Informal Sectors • Shadow Economy • Institutional Arrangements | development issues related to formal and informal sectors, shadow economy, and legal, social, economic, and political institutional arrangements, including privatization, property rights, and social unrest. | institutions, rights, human rights, human |
| | P48 | Political Economy • Legal Institutions • Property Rights • Natural Resources • Energy • Environment • Regional Studies | issues related to political economy, legal institutions and property rights in economic systems other than capitalist, socialist, and transitional economic systems. This includes roles of government and/or power relationships in resource allocation. | institutions, rights |
| | F33 | International Monetary Arrangements and Institutions | issues related to exchange rate regimes, including fixed or flexible exchange rate systems and their impacts. Also covers studies related to international financial organizations, for example the IMF or the World Bank, and their roles. | institutions, rights |
| Goal 17. Strengthen the means of implementation and revitalize the global partnership for sustainable development | F63 | Economic Development | the impact of globalization on economic development, including poverty, labor markets, and gender issues. | developed, development, developing |
| | O11 | Macroeconomic Analyses of Economic Development | macroeconomic issues related to development, including economic growth of developing countries. These studies include both theoretical models and empirical studies (time series and cross-sectional). | development, data, developing |
| | I25 | Education and Economic Development | the impact of education on economic development, and the impact of levels of development on education. May cross-list with O15. | development, developing, capacity |

Notes: * refers to cases where there are also other JEL codes with equal keyword overlap and the choice of this specific JEL code among the top three is based on discretion. Ranking of the top3 JEL codes is based on the following weighting:  weight 1 for the first five keywords and 1/[number of order] for the rest. Data available from the author upon request.

## Table 5. Link between SDGs and JEL codes, selected three keywords

| SDG | Top three JEL codes | | Covers studies about/related to | Overlapping keywords |
|---|---|---|---|---|
| Goal 1. End poverty in all its forms everywhere | I32 | Measurement and Analysis of Poverty | issues related to poverty, its measurement and analysis. | poverty, poor |
| | O15* | Human Resources • Human Development • Income Distribution • Migration | issues related to labor, demography, health, education, and welfare in the context of economic development and developing economies. Includes studies on nutrition, health, education, fertility, household structure and formation, labor markets and social policy. | poverty |
| | H53* | Government Expenditures and Welfare Programs | issues related to government expenditures on welfare and related policies, including food stamp programs and studies about HUD. | poverty |
| Goal 2. End hunger, achieve food security and improved nutrition and promote sustainable agriculture | O13 | Agriculture • Natural Resources • Energy • Environment • Other Primary Products | development issues related to agriculture, natural resources, other primary products, energy, and the environment. | food, agriculture |
| | Q11 | Aggregate Supply and Demand Analysis • Prices | issues related to aggregate agricultural market, including supply, demand, prices and sustainable agriculture. | food, agriculture |
| | Q10 | General | general economic issues related to agriculture, including survey articles, textbooks, and data. | food, agriculture |
| Goal 3. Ensure healthy lives and promote well-being for all at all ages | I12 | Health Behavior | issues related to health production and conditions and their consequences. Covers studies about the causes and effects of being healthy or unhealthy, including studies about quality-adjusted life years. Includes studies on nutrition, mortality, morbidity, suicide, substance abuse and addiction, and disability as related to economic behavior. Studies should be cross-classified here and under I11 if they cover topics in both categories. | health, disease, mortality |
| | I15 | Health and Economic Development | the impact of health on economic development, and the impact of levels of development on health. May cross-list with O15. | health, disease, mortality |
| | I14 | Health and Inequality | the impact of health on social and economic inequality, and the impact of inequality on health. May cross-list with O15. | health, mortality |
| Goal 4. Ensure inclusive and equitable quality education and promote lifelong learning opportunities for all | I21 | Analysis of Education | all the economic issues, including output, quality, and demand and supply, related to education, except for educational finance. | education, school, learning |
| | J24 | Human Capital • Skills • Occupational Choice • Labor Productivity | issues related to labor quality. Includes studies about the effects of education as human capital (or the pecuniary return to schooling) as well as those about training programs (public or private) and occupational choices. Micro-studies about labor productivity are also classified here. | education, school, learning |
| | A21* | Pre-college | issues related to teaching economics at a pre-college level. For example, it includes studies on how to teach economics, what to teach in economics, and the outcomes from economic education at the pre-college level. Also covers the economic knowledge of the public at large. | education, school |
| Goal 5. Achieve gender equality and empower all women and girls | J16 | Economics of Gender • Non-labor Discrimination | economic issues related to gender, except for labor market discrimination, which is classified under J71 or J78. | women, gender, equality |
| | J71 | Discrimination | issues related to labor discrimination, including wage discrimination and discrimination in hiring and firing. | women, gender, equality |
| | F63* | Economic Development | the impact of globalization on economic development, including poverty, labor markets, and gender issues. | gender, equality |
| Goal 6. Ensure availability and sustainable management of water and sanitation for all | Q53* | Air Pollution • Water Pollution • Noise • Hazardous Waste • Solid Waste • Recycling | issues related to environmental degradation, its impacts and its solution. | water |
| | Q25* | Water | issues related to water as a renewable resource and its conservation. | water |
| | L95* | Gas Utilities • Pipelines • Water Utilities | issues related to the gas and water utility industries, including studies about pipelines. | water |
| Goal 7. Ensure access to affordable, reliable, sustainable and modern energy for all | Q42 | Alternative Energy Sources | issues related to alternative energy sources, for example non-hydrocarbon fuels or wind. | energy, renewal |
| | L94 | Electric Utilities | issues related to the electric utility industry. | energy, electricity |
| | Q48* | Government Policy | issues related to government policy on energy. | energy |
| Goal 8. Promote sustained, inclusive and sustainable economic growth, full and productive employment and decent work for all | E24 | Employment • Unemployment • Wages • Intergenerational Income Distribution • Aggregate Human Capital • Aggregate Labor Productivity | issues related to the aggregate labor market, including unemployment, wage level, wage indexation, intergenerational income distribution and aggregate labor productivity. Also covers studies about intergenerational income mobility. | labor, employment, productivity |
| | J24 | Human Capital • Skills • Occupational Choice • Labor Productivity | issues related to labor quality. Includes studies about the effects of education as human capital (or the pecuniary return to schooling) as well as those about training programs (public or private) and occupational choices. Micro-studies about labor productivity are also classified here. | labor, employment, productivity |
| | J68* | Public Policy | public policy issues related to labor mobility and unemployment, including employment services. | labor, employment |
| Goal 9. Build resilient infrastructure, promote inclusive and sustainable industrialization and foster innovation | H54* | Infrastructures • Other Public Investment and Capital Stock | issues related to national expenditures on infrastructures and other public investment and their impacts on the performance of the national or local economy. | infrastructure |
| | R42* | Government and Private Investment Analysis • Road Maintenance • Transportation Planning | issues related to investment in transportation and transportation infrastructure. | infrastructure |
| | H52* | Government Expenditures and Education | issues related to government expenditures on education and related policies. | infrastructure |

## Table 5. Continued

| | | | | |
|---|---|---|---|---|
| Goal 10. Reduce inequality within and among countries | D31* | Personal Income, Wealth, and Their Distributions | issues related to personal income and wealth and their distributions, most of which are empirical. Studies about wage as a component of personal income should be classified here. | inequality, income |
| | F61* | Microeconomic Impacts | to the impact of globalization on various microeconomic issues, including income distribution, market structure and pricing, and the behavior of consumers and firms. | inequality, income |
| | I24* | Education and Inequality | the impact of education on social and economic inequality, and the impact of inequality on education. May cross-list with O15. | inequality, income |
| Goal 11. Make cities and human settlements inclusive, safe, resilient and sustainable | O18 | Urban, Rural, Regional, and Transportation Analysis • Housing • Infrastructure | regional, urban, rural, and transportation issues related to economic development and developing economies, including urbanization. | cities, urban |
| | R11* | Regional Economic Activity: Growth, Development, Environmental Issues, and Changes | issues related to regional economic activity including growth, development, and changes. Studies may also include the effects of agglomeration economies and foreign linkages (for example foreign direct investment or foreign trade) on regional economic activities. Includes environmental aspects of regional economics, such as the regional effects of climate change. | cities, urban |
| | R12* | Size and Spatial Distributions of Regional Economic Activity | issues related to spatial distributions of economic activity including economic geography, interregional trade, and regional convergence. | urban, housing |
| Goal 12. Ensure sustainable consumption and production patterns | D62 | Externalities | mostly theoretical studies about issues related to externalities and welfare analysis. | consumption, production |
| | E27 | Forecasting and Simulation: Models and Applications | issues related to forecasting or simulation of aggregate consumption, saving, production, employment and/or investment. | consumption, production |
| | Q11 | Aggregate Supply and Demand Analysis • Prices | issues related to aggregate agricultural market, including supply, demand, prices and sustainable agriculture. | production, sustainable |
| Goal 13. Take urgent action to combat climate change and its impacts* | Q54 | Climate • Natural Disasters and Their Management • Global Warming | issues related to climate and natural disasters, including desertification and drought. | climate, climate change, emission |
| | Q15 | Land Ownership and Tenure • Land Reform • Land Use • Irrigation • Agriculture and Environment | issues related to agricultural land, including ownership and tenure (for example, issues related to share cropping), irrigation and agricultural land valuation. | climate, emission |
| | Q58 | Government Policy | issues related to both national and international government policy pertaining to the environment, including environmental taxes and tradable permits. | climate, emission |
| Goal 14. Conserve and sustainably use the oceans, seas and marine resources for sustainable development | L92 | Railroads and Other Surface Transportation | issues related to surface transportation, including trains, autos, buses, trucks, and water carries. Also covers studies related to ports as they influence the performance of the transportation industry. | marine, ocean |
| | Q25 | Water | issues related to water as a renewable resource and its conservation. | ocean |
| | Q22* | Fishery • Aquaculture | issues related to fishery and aquaculture. | fish |
| Goal 15. Protect, restore and promote sustainable use of terrestrial ecosystems, sustainably manage forests, combat desertification, and halt and reverse land degradation and halt biodiversity loss | O13 | Agriculture • Natural Resources • Energy • Environment • Other Primary Products | development issues related to agriculture, natural resources, other primary products, energy, and the environment. | land, forest |
| | Q23 | Forestry | issues related to forestry. | land, forest |
| | Q57* | Ecological Economics: Ecosystem Services • Biodiversity Conservation • Bioeconomics • Industrial Ecology | issues related to topics in ecological economics including ecosystem services and biodiversity. | biodiversity |
| Goal 16. Promote peaceful and inclusive societies for sustainable development, provide access to justice for all and build effective, accountable and inclusive institutions at all levels | O17 | Formal and Informal Sectors • Shadow Economy • Institutional Arrangements | development issues related to formal and informal sectors, shadow economy, and legal, social, economic, and political institutional arrangements, including privatization, property rights, and social unrest. | institutions, human rights |
| | P48* | Political Economy • Legal Institutions • Property Rights • Natural Resources • Energy • Environment • Regional Studies | issues related to political economy, legal institutions and property rights in economic systems other than capitalist, socialist, and transitional economic systems. This includes roles of government and/or power relationships in resource allocation. | institutions |
| | F33* | International Monetary Arrangements and Institutions | issues related to exchange rate regimes, including fixed or flexible exchange rate systems and their impacts. Also covers studies related to international financial organizations, for example the IMF or the World Bank, and their roles. | institutions |
| Goal 17. Strengthen the means of implementation and revitalize the global partnership for sustainable development | F35 | Foreign Aid | issues related to foreign aid in general. | development, official development assistance |
| | O19 | International Linkages to Development • Role of International Organizations | issues related to international linkages to development, including the role of international organizations, foreign aid, trade and financial flows, and topics such as IMF conditionality. | development, trade |
| | Q56 | Environment and Development • Environment and Trade • Sustainability • Environmental Accounts and Accounting • Environmental Equity • Population Growth | issues related to the environment and development, the environment and trade, environmental accountability, environmental equity, and environmental sustainability. | development, trade |

Notes: * refers to cases where there are also other JEL codes with equal keyword overlap and the choice of this specific JEL code among the top three is based on discretion. Data available from the author upon request.

The mappings presented in Tables 3, 4 and 5 are examples of very simple ways to link JEL codes and SDGs and there are naturally multiple other ways. In other words, they are necessarily imperfect and incomplete and have several limitations. It is worth noting that some JEL codes cannot be directly linked to specific SDGs. For instance, economic methods (e.g., JEL code class C "Mathematical and Quantitative Methods" with its subclasses) are "general purpose technologies" and are not SDG-specific but can be applied to study all of them. It should also be noted that for some JEL codes the guideline mentions that documents classified to that specific JEL code may, or in some cases should, be "cross-listed" or "cross-classified" to other JEL codes. For instance, the guideline for JEL code "I15 Health and Economic Development" notes that it "covers studies about the impact of health on economic development, and the impact of levels of development on health. May cross-list with O15" (see Goal 3 in Table 4). Similarly, the guideline for JEL code "J83 Workers' Rights" notes that it "covers studies about labor standard issues related to workers' rights. Studies should be cross-classified here and under the other appropriate categories in J, if they are relevant." These nuances in JEL- code-specific guidelines were ignored in this article.

## 4 Discussion

As the sustainable development goals lack their corresponding JEL codes, this paper has suggested simple ways to connect SDGs and JEL codes. There are multiple ways to improve these linkages. For instance, one could use the corpus or keywords of economics articles that are tagged with a specific JEL code to establish the link to the closest SDGs. This would require much deeper bibliometric, topic modelling and science mapping analyses (cf. Aria & Cuccurullo 2017; Körfgen et al. 2018; Ambrosino et al. 2019; LaFleur 2019). These are left for future studies.

The correspondence between economic research topics and SDGs illustrates that economics is not after all so dismal science as some may think. Highlighting the fact that economists study research topics that are closely related to SDGs may encourage other researchers to allocate more attention on analysing how economic research can help in solving societal challenges and reaching all the SDGs by 2030.